\documentclass[pre,aps,twocolumn,superscriptaddress,nofootinbib]{revtex4-1}
\usepackage{amssymb}
\usepackage{amsmath}
\usepackage{subfigure}
\usepackage{graphicx}
\usepackage{textcomp}
\usepackage{url}
\usepackage{color}
\usepackage{cases}
\usepackage[normalem]{ulem}

\usepackage[colorlinks=true, urlcolor=blue, anchorcolor=blue, citecolor=blue,filecolor=blue,linkcolor=blue,menucolor=blue
]{hyperref}

\usepackage[titletoc,title]{appendix}

\usepackage{color}

\begin{document}
\title{Large deviations of the interface height in the Golubovi\'{c}-Bruinsma model of stochastic growth}

\author{Baruch Meerson}
\email{meerson@mail.huji.ac.il}
\author{Arkady Vilenkin}
\email{vilenkin@mail.huji.ac.il}
\affiliation{Racah Institute of Physics, Hebrew University of
Jerusalem, Jerusalem 91904, Israel}


\begin{abstract}
We study large deviations of the one-point height distribution, $\mathcal{P}(H,T)$, of a stochastic interface, governed by the Golubovi\'{c}-Bruinsma equation
$$
\partial_{t}h=-\nu\partial_{x}^{4}h+\frac{\lambda}{2}\left(\partial_{x}h\right)^{2}+\sqrt{D}\,\xi(x,t)\,,
$$
where $h(x,t)$ is the interface height at point $x$ and time $t$, and $\xi(x,t)$ is the Gaussian white noise. The interface is initially flat, and $H$ is defined by the relation $h(x=0,t=T)=H$. Using the optimal fluctuation method (OFM), we focus on the short-time limit. Here the typical fluctuations of $H$ are Gaussian, and we evaluate the strongly asymmetric and non-Gaussian tails of $\mathcal{P}(H,T)$. We show that the upper tail scales as $-\ln \mathcal{P}(H,T) \sim H^{11/6}/T^{5/6}$. The lower tail, which scales as $-\ln \mathcal{P}(H,T) \sim H^{5/2}/T^{1/2}$, coincides with its counterpart for the Kardar-Parisi-Zhang equation, and we uncover a simple physical mechanism behind this universality.  Finally, we verify our asymptotic results for the tails, and compute the large deviation function of $H$, numerically.

\end{abstract}

\maketitle

\section{Introduction}

It is natural to start our story with the celebrated Kardar-Parisi-Zhang (KPZ) equation: a paradigmatic model of non-equilibrium stochastic growth. In one dimension, this equation has the form \cite{KPZ}
\begin{equation}
\label{eq:KPZ}
\partial_{t}h=\nu\partial_{x}^{2}h+\frac{\lambda}{2}\left(\partial_{x}h\right)^{2}+\sqrt{D}\,\xi(x,t)\,,
\end{equation}
where $h(x,t)$ is the height of a growing KPZ interface at the point $x$ of a substrate at time $t$,
and $\xi(x,t)$ is a Gaussian white noise with zero average
and
\begin{equation}\label{correlator}
\langle\xi(x_{1},t_{1})\xi(x_{2},t_{2})\rangle = \delta(x_{1}-x_{2})\delta(t_{1}-t_{2}).
\end{equation}
At late times, the lateral correlation length of the one-dimensional KPZ interface grows as $t^{2/3}$, and the characteristic interface width grows as $t^{1/3}$. The exponents $2/3$ and $1/3$ are hallmarks of the KPZ universality class: an important universality class of non-equilibrium growth \cite{Vicsek,HHZ,Barabasi,Krug1997,Corwin,QS,S2016,Takeuchi2017}.

In the last decade, more detailed characteristics of the height fluctuations of the KPZ interface have been introduced and studied. One of these characteristics is the probability distribution $\mathcal{P}\left(H,t\right)$ of the interface height at specified point and time, $H=h\left(x=0,t\right)$. This distribution strongly depends on the initial condition $h\left(x,t=0\right)$. Remarkably, in an infinite system the dependence on the initial conditions persists forever \cite{QS,S2016,Takeuchi2017}. There has been a spectacular progress in quantitative analysis of this problem. It started from the discovery of remarkable exact representations for $\mathcal{P}\left(H,t\right)$ for several
``standard" initial conditions, see Refs. \cite{QS,S2016,Takeuchi2017} for reviews. This line of work received the name ``stochastic integrability". The progress continued in the form of application to the KPZ equation of the optimal fluctuation method (OFM)
\cite{KK2007,KK2008,KK2009,MKV,KMSparabola,Janas2016,SMS2017,MeersonSchmidt2017,SMS2018,SKM2018,SmithMeerson2018,MV2018,Asida2019,
SMV2019,HMS2019,HMS2021}. It was found that the OFM captures the whole large-deviation function of $H$ at short times (and correctly describes far distribution tails of at all times) for a broad variety of initial and boundary conditions. The crux of the OFM
is the determination of the optimal path, that is the most likely history of the interface and the most likely realization of the noise which dominate the contribution of different histories to $\mathcal{P}(H,T)$ at specified $H$. The optimal paths, determined in Refs. \cite{KK2007,KK2008,KK2009,MKV,KMSparabola,Janas2016,SMS2017,MeersonSchmidt2017,SMS2018,SKM2018,SmithMeerson2018,MV2018,Asida2019,
SMV2019,HMS2019,HMS2021,Smith2022}, provided an instructive (and often fascinating) insight into physics of large deviations of the KPZ interface height. Some of the optimal paths were also directly measured in Monte Carlo simulations which probe the tails with an importance sampling algorithm \cite{HMS2019,HMS2021}.

In the last two years the subject has received a renewed attention. It was recognized some time ago \cite{Janas2016} that the OFM equations for Eq.~(\ref{eq:KPZ}) belong to a class of completely integrable classical systems. Recently this idea has borne fruit in Refs. \cite{KLD1,KLD2} where the short-time large-deviation function of $H$ was calculated exactly, for three major initial conditions, by the inverse scattering method.

The famous KPZ equation, however, is only one of a whole family of continuum models of noneqilibrium stochastic interface growth. Since the beginning of the nineties of the last century, several other models, with different mechanisms of nonlinearity and relaxation and different properties of noise, have been proposed and studied, see the book \cite{Barabasi} for an extensive review. We believe that the time is ripe to broaden the application range of the OFM by applying it to some of these systems. Here we focus on the Golubovi\'{c}-Bruinsma (GB) equation of stochastic interface growth \cite{GB}:
\begin{equation}
\label{GBdimensional}
\partial_{t}h=-\nu\partial_{x}^{4}h+\frac{\lambda}{2}\left(\partial_{x}h\right)^{2}+\sqrt{D}\,\xi(x,t)\,.
\end{equation}
The GB equation is a model equation, which keeps the same nonlinearity and noise as in the KPZ equation (\ref{eq:KPZ}), but differs from the latter by the relaxation mechanism: here it is surface diffusion, described by the fourth-derivative term $-\nu\partial_{x}^{4}h$. As we shall see, this difference can be quite important.

Without the nonlinear term,  the GB equation (\ref{GBdimensional}) becomes the linear stochastic Mullins-Herring equation:
\begin{equation}
\label{MH}
\partial_{t}h=-\nu\partial_{x}^{4}h+\sqrt{D}\,\xi(x,t)\,.
\end{equation}
Its noiseless version $\partial_t h=-\nu\partial_{x}^{4}h$, proposed by Mullins
65 years ago \cite{Mullins}, describes the capillary relaxation
of a solid surface, where
the surface diffusion of adatoms is accompanied by adatom
exchange between the surface and the bulk of the solid \cite{WV,Villain,Luse,Taylor,VB}. The linear stochastic equation~(\ref{MH})
was introduced by Wolf and Villain \cite{WV} and by Das Sarma and Taborenea \cite{DST}. This equation and its analog with a conserved noise were extensively studied in the context of dynamic scaling
behavior of the interface \cite{Racz,Siegert,DST,Barabasi,Krug1997}. Previously  we employed the OFM to study the probability distribution $\mathcal{P}(H,T)$ of the one-point interface height for Eq.~(\ref{MH}) with a conserved noise \cite{MV2016}. Here we
study the short-time behavior of $\mathcal{P}(H,T)$  for the nonlinear GB equation (\ref{GBdimensional}). We suppose that the process starts at $t=0$ from flat interface,
\begin{equation}\label{IC}
  h(x,t=0)=0\,,
\end{equation}
and condition the interface height at $x=0$ on reaching a specified value $H$ at specified time $t=T$ \cite{shift}:
\begin{equation}\label{H}
h\left(x=0,t=T\right) =H \,.
\end{equation}
In the short-time limit, $T \ll T_{\text{NL}}$ (where $T_{\text{NL}} = \nu^{5/7} (D\lambda^2)^{-4/7}$ is the characteristic nonlinear time of the GB equation),  typical fluctuations of $H$ are governed by the linear equation (\ref{MH}). Large deviations of $H$, however, ``feel" the nonlinear term in the GB equation from the start, and they demand a full account of the nonlinearity.

It is convenient for the following to rescale the variables. Upon the rescaling transformation
\begin{equation}\label{rescaling}
\frac{t}{T} \to t,\quad \frac{x}{(\nu T)^{1/4}} \to x\quad\text{and}\quad \frac{|\lambda| T^{1/2}h}{\nu^{1/2}}\to h
\end{equation}
Eq. (\ref{GBdimensional}) becomes
\begin{equation}
\label{GB}
\partial_{t}h=-\partial_{x}^{4}h-\frac{1}{2}\left(\partial_{x}h\right)^{2}+\sqrt{\epsilon}\,\xi(x,t)\,,
\end{equation}
where $\epsilon = D\lambda^2 T^{7/4}/\nu^{5/4}$, and we have assumed, without loss of generality, that $\lambda<0$.
Clearly, $\mathcal{P}\left(H,T\right)$ depends only on two dimensionless parameters: the rescaled height $\tilde{H}=|\lambda| T^{1/2} H/\nu^{1/2}$ and $\epsilon$.

As we already stated, typical fluctuations, $|H|\ll 1$, at short times are Gaussian. Using the OFM, we will obtain
\begin{equation}\label{Gauss}
-\ln \mathcal{P} (H,T)\simeq  \frac{3 \pi \nu^{1/4}H^2}{2^{3/4} \Gamma(1/4) D T^{3/4}}\,,
\end{equation}
where $\Gamma(\dots)$ is the gamma function. As to be expected, this result is in perfect agreement with previous results  for the linear equation~(\ref{MH}) \cite{Barabasi,Krug1997,SMS2017}.

The tails of  $\mathcal{P}\left(H,T\right)$ are non-Gaussian. For the upper tail $\lambda H \to \infty$  we obtain a slower-than-Gaussian asymptotic
\begin{equation}\label{11over6}
-\ln \mathcal{P}(H,T) \simeq \frac{\beta \nu^{1/3} H^{11/6}}{D |\lambda|^{11/6}T^{5/6}}\,,
\end{equation}
where $\beta\simeq 1.73$. The asymptotic~(\ref{11over6}) is quite different from the  $\lambda H \to \infty$  asymptotic of the one-point height distribution for the KPZ equation. The latter scales as $-\ln \mathcal{P}(H,T) \sim H^{3/2}/T^{1/2}$.

The lower tail $\lambda H \to -\infty$  is controlled by the noise and nonlinearity, and independent of the surface diffusion:
\begin{equation}\label{5over2}
-\ln \mathcal{P}(H,T) \simeq \frac{8\sqrt{2 |\lambda|} \, H^{5/2}}{15 \pi D T^{1/2}}\,.
\end{equation}
This faster-than-Gaussian tail coincides  with the leading-order $\lambda H \to -\infty$ tail of $\mathcal{P}(H,T)$ for the KPZ equation, and we point out to a simple physical mechanism behind this universality.

Here is a plan of the remainder of the paper.  In Sec. \ref{OFM} we present the OFM equations and boundary conditions for the evaluation of $\mathcal{P}(H,T)$. Section \ref{OFM}  also includes our numerical results for the complete large-deviation function of $H$, obtained with a back-and-forth iteration algorithm \cite{Chernykh}. In Sec. \ref{solving} we present asymptotic solutions of the OFM problem in three different limits: $|H|\ll 1$ (typical fluctuations), $\lambda H \to \infty$ (the upper tail), and $\lambda H \to -\infty$ (the lower tail). These asymptotic solutions are motivated by our numerical results for the optimal paths in different regimes, and by expected analogies with the OFM results for the KPZ equation. Finally, Sec. \ref{discussion} contains a brief summary and a discussion of possible extensions of our results.

\section{OFM formalism}
\label{OFM}

The OFM (also known as the weak-noise theory, the instanton method, the dissipative WKB approximation, the macroscopic fluctuation theory, \textit{etc}.) is a widely used asymptotic method based on a saddle-point evaluation of the exact path integral for the GB equation (\ref{GB}). The method relies on a small parameter (in our case one formally sets $\epsilon \to 0$).  In this asymptotic regime the weak-noise scaling holds:
\begin{equation}\label{lnP}
-\ln \mathcal{P} (H,T)\simeq \frac{s (\tilde{H})}{\epsilon}= \frac{\nu^{5/4}}{D \lambda^2 T^{7/4}}\,s\left(\frac{|\lambda| T^{1/2}H}{\nu^{1/2}}\right)\,,
\end{equation}
and we will suppress the tilde in $\tilde{H}$ in the following. The \textit{a priori} unknown large deviation function $s(\dots)$ comes  from the solution of the minimization problem, intrinsic in the saddle-point evaluation. The minimization problem can be recast as an effective classical field theory, the solution of which obeys the Hamilton's equations
of the OFM:
\begin{eqnarray}
  \partial_{t} h &=& -\partial_{x}^4 h -(1/2) \left(\partial_x h\right)^2+\rho\,,   \label{eqh}\\
  \partial_{t}\rho &=& \partial_{x}^4 \rho - \partial_x \left(\rho \partial_x h\right)\,. \label{eqrho}
\end{eqnarray}
These equations describe the optimal path, that is the most likely histories of the interface $h(x,t)$ and of the rescaled noise $\rho(x,t)$, conditioned on $H$.
The interface is initially flat, see Eq. (\ref{IC}). The ``final" condition  (\ref{H}) can be temporarily incorporated with the help of a Lagrange multiplier which leads to the  boundary condition
\begin{equation}\label{pT}
    \rho(x,1)=\Lambda \,\delta(x).
\end{equation}
The \textit{a priori} unknown Lagrange multiplier $\Lambda$ is ultimately expressed through $H$. The derivation of
Eqs.~(\ref{eqh})-(\ref{pT}) closely follows that for the KPZ equation (\ref{eq:KPZ}), see \textit{e.g.} Ref. \cite{MKV}.

Once the optimal path is determined, the large deviation function of $H$ -- that is, the rescaled action $s(H)$ -- can be calculated from the equation
\begin{equation}\label{action}
s= \frac{1}{2}\int_0^1 dt \int_{-\infty}^{\infty} dx\,\rho^2(x,t)\,.
\end{equation}
Alternatively, $s(H)$ can be found from the ``shortcut relation" $ds/dH =\Lambda$, see \textit{e.g.} Ref. \cite{Vivo}.

It is often convenient to rewrite Eqs.~(\ref{eqh}) and (\ref{eqrho}) in terms of the interface slope
$V(x,t) = \partial_x h(x,t)$:
\begin{eqnarray}
  \partial_{t} V +V \partial_x V &=& -\partial_{x}^4 V +\partial_x\rho\,,   \label{eqV}\\
  \partial_{t}\rho + \partial_x \left(\rho V\right)&=& \partial_{x}^4 \rho \,. \label{eqrhoV}
\end{eqnarray}

\begin{figure}
\includegraphics[width=0.30\textwidth,clip=]{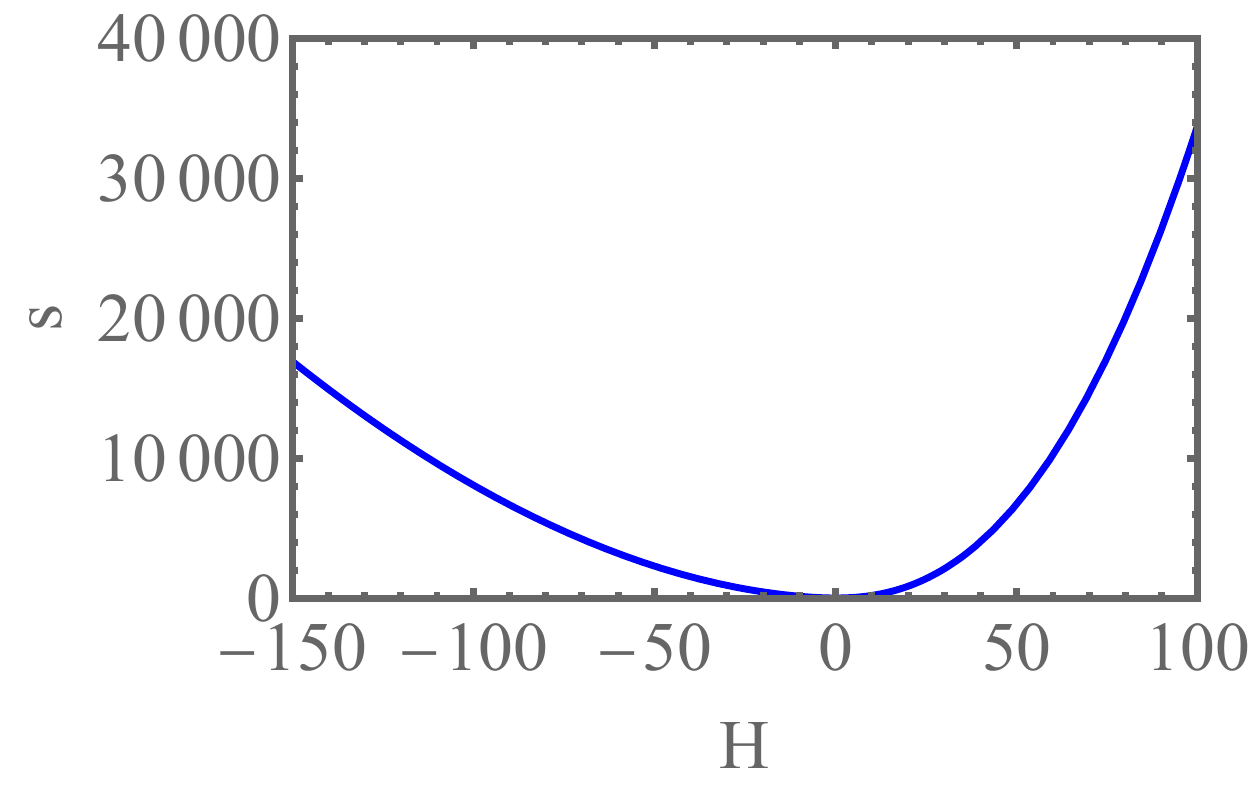}
\caption{The large-deviation function $s(H)$ computed numerically. Evident is a strong asymmetry of the tails.}
\label{sHnum}
\end{figure}

Figure \ref{sHnum} shows the rescaled action $s(H)$,  obtained by solving the OFM problem numerically with a modified back-and-forth iteration algorithm \cite{Chernykh,numerics}. Evident in Fig. \ref{sHnum} is a strong asymmetry of the right and left tails of $\mathcal{P}(H,T)$.

\section{Asymptotics of $s(H)$}
\label{solving}

\subsection{Typical fluctuations, $|H|\ll 1$.}
\begin{figure}
\includegraphics[width=0.30\textwidth,clip=]{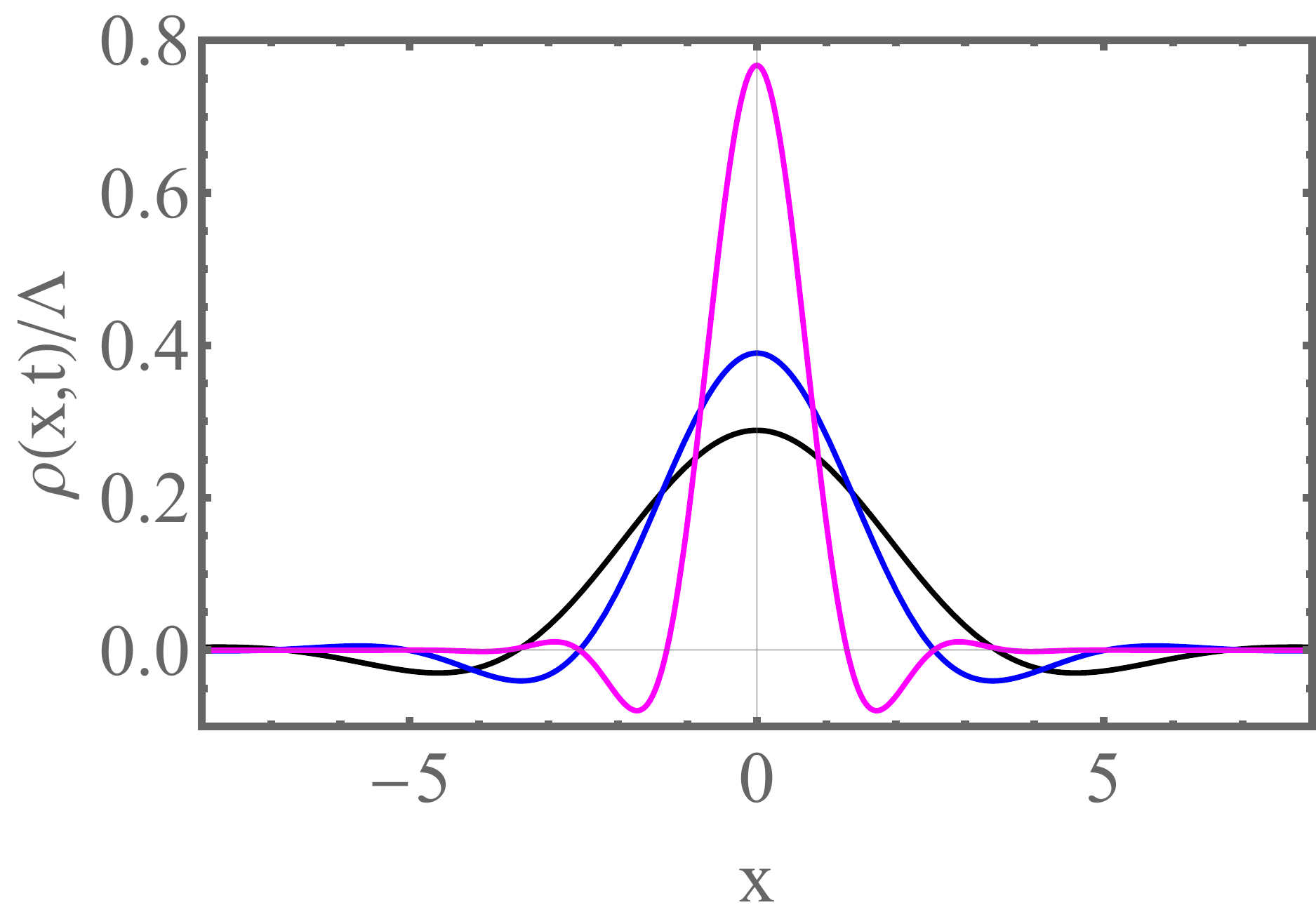}
\caption{The optimal history of rescaled noise $\rho(x,t)/\Lambda$ in the limit of small fluctuations, $|H|\ll 1$.  Shown is the solution~(\ref{Green}) at $t=0$ (black), $0.7$ (blue) and $0.85$ (magenta).}
\label{figlin}
\end{figure}

Typical, small fluctuations of $H$ are described by the limit of $H\to 0$ or $\Lambda \to 0$. In this limit we can linearize Eqs.~(\ref{eqh}) and  (\ref{eqrho}) and obtain:
\begin{eqnarray}
  \partial_{t} h &=& -\partial_{x}^4 h +\rho\,,  \label{eqhlin}\\
  \partial_{t}\rho &=& \partial_{x}^4 \rho\,.\label{eqrholin}
\end{eqnarray}
These linear equations provide the optimal fluctuation theory for Eq.~(\ref{MH}), and they can be solved exactly. Solving Eq.~(\ref{eqrholin}) backward in time with the initial condition (\ref{pT}), we obtain
\begin{equation}\label{Green}
p(x,t)=\frac{\Lambda}{(1-t)^{1/4}} \Phi \left[\frac{x}{(1-t)^{1/4}}\right]\,,
\end{equation}
where
\begin{eqnarray}
 \Phi(z) &=& \frac{\Gamma (5/4)}{\pi}\, _0F_2\left(\frac{1}{2},\frac{3}{4};\frac{z^4}{2
   56}\right) \nonumber \\
         &-& \frac{\Gamma(3/4)}{8\pi} z^2 \,_0F_2\left(\frac{5}{4},\frac{3}{2};\frac{z^4}{2
   56}\right)\,.  \label{Phi}
\end{eqnarray}
Here $_pF_q(a;b;z)$ is the generalized hypergeometric function \cite{Wolfram1}. This solution at different times is depicted on Fig. \ref{figlin}. It exhibits an oscillatory spatial decay, characteristic of the surface diffusion \cite{Mullins}.

\begin{figure}
\includegraphics[width=0.30\textwidth,clip=]{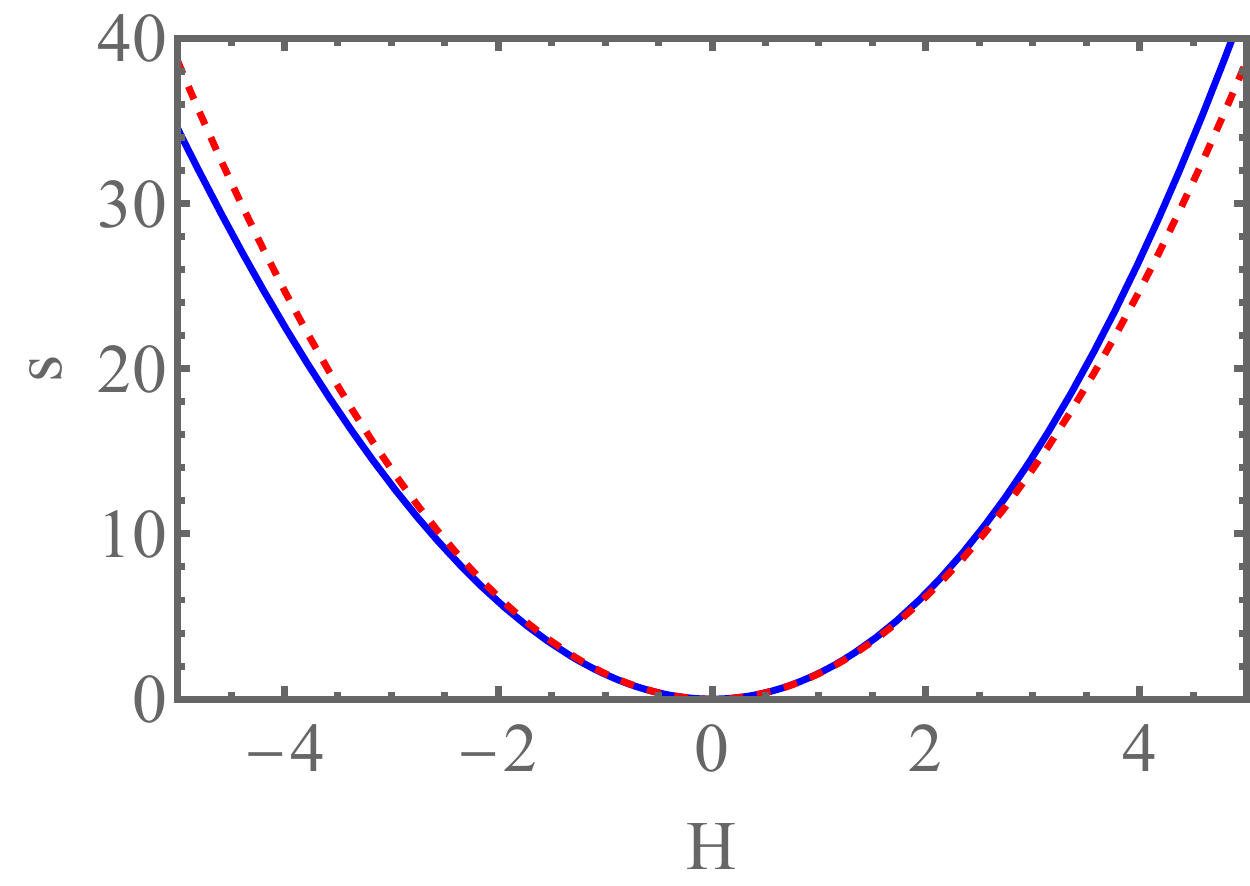}
\caption{The quadratic asymptotic ~(\ref{slinear2}) of $s(H)$ (dashed line) versus $s(H)$ found numerically (solid line).}
\label{sHG}
\end{figure}

The action in terms of $\Lambda$ can be obtained by plugging Eqs.~(\ref{Green}) and~(\ref{Phi}) into Eq.~(\ref{action}):
\begin{equation}\label{slin1}
s(\Lambda) = \frac{\Lambda^2}{2}\int_0^1\frac{dt}{(1-t)^{1/4}}\int_{-\infty}^{\infty} dz\,\Phi^2(z) = \alpha \Lambda^2\,,
\end{equation}
where $\alpha = \frac{\Gamma \left(1/4\right)}{6\sqrt[4]{2} \pi} =0.16174\dots$.  Now we have to express $\Lambda$ through $H$. We can bypass
the need to solve Eq.~(\ref{eqhlin}) by using ``the shortcut relation" $ds/dH =\Lambda$.  We have
\begin{equation}\label{dsdH}
\frac{ds}{dH} = \frac{ds}{d\Lambda} \frac{d\Lambda}{dH} = 2 \alpha \Lambda \frac{d\Lambda}{dH}=\Lambda\,.
\end{equation}
Therefore $\Lambda=H/2\alpha$, and we obtain
\begin{equation}\label{slinear2}
s(|H|\ll 1) \simeq \frac{H^2}{4 \alpha}\,,
\end{equation}
leading to  Eq.~(\ref{Gauss}) for $\mathcal{P}(H,T)$. Figure \ref{sHG} compares the asymptotic~(\ref{slinear2}) with the numerically computed $s(H)$, and a good agreement is observed for small $|H|$. The Gaussian asymptotic (\ref{Gauss}), which predicts, in the original variables, a $T^{3/8}$ growth of the standard deviation of $H$ with time, perfectly agrees with the known results for Eq.~(\ref{MH}), see page 142 of Ref. \cite{Barabasi}, Eq. (3.25) of Ref. \cite{Krug1997} and Eq. (8) of Ref. \cite{SMS2017}. Notice, that the applicability condition of this asymptotic
$|H|\ll 1$ in the rescaled variables, depends on time $T$ in the original variables:
$$
|H|\ll \frac{\nu^{1/2}}{|\lambda| T^{1/2}}\,.
$$
Therefore, as $T\to 0$ at fixed $H$, the Gaussian asymptotic is always valid. However, for any fixed $T$, this asymptotic breaks down for sufficiently large $|H|$, that is in the distribution tails.   Let us now describe these non-Gaussian tails.

\subsection{Upper tail: $\lambda H \to \infty$}

\begin{figure}
\includegraphics[width=0.30\textwidth,clip=]{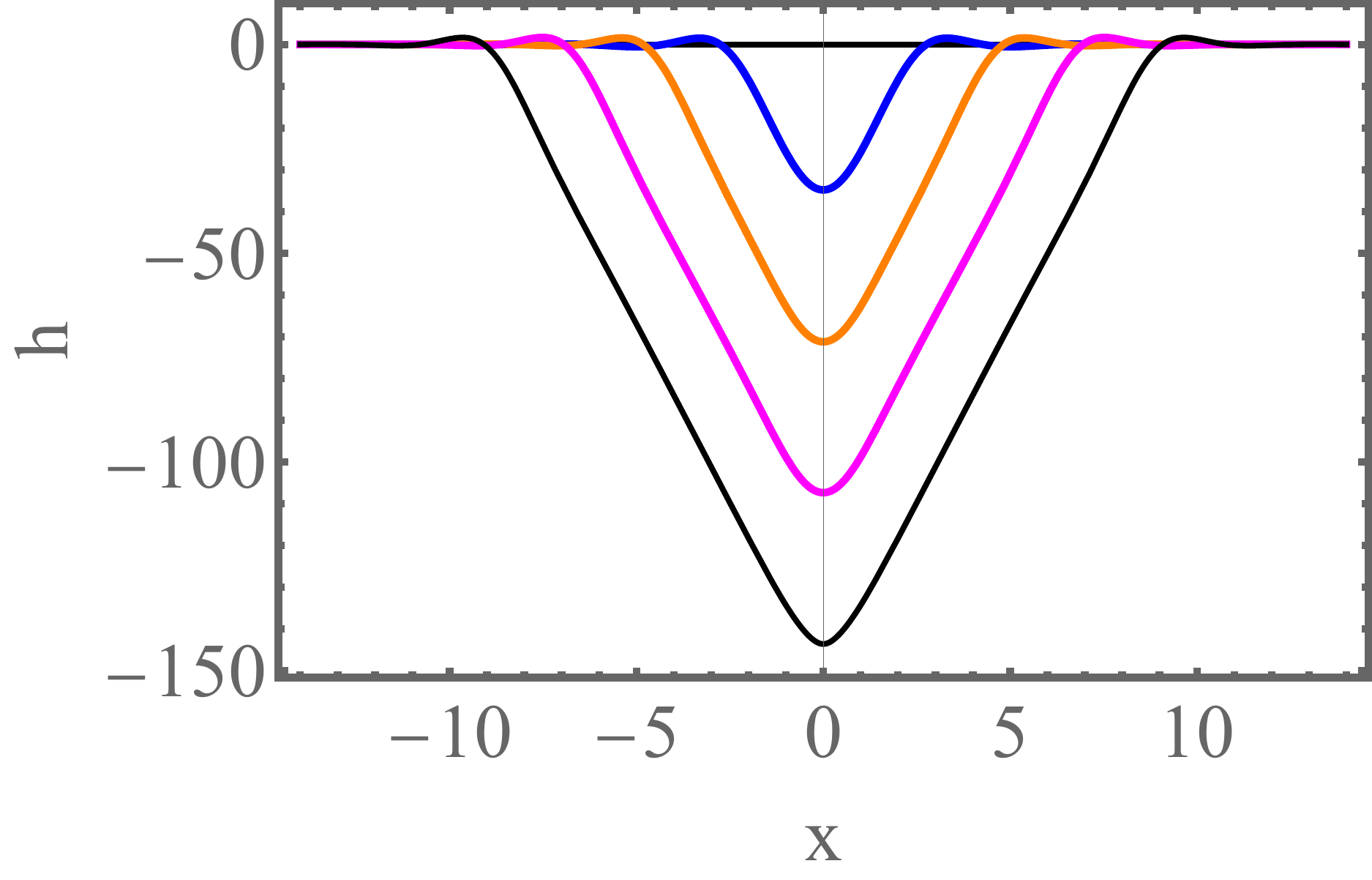}
\includegraphics[width=0.30\textwidth,clip=]{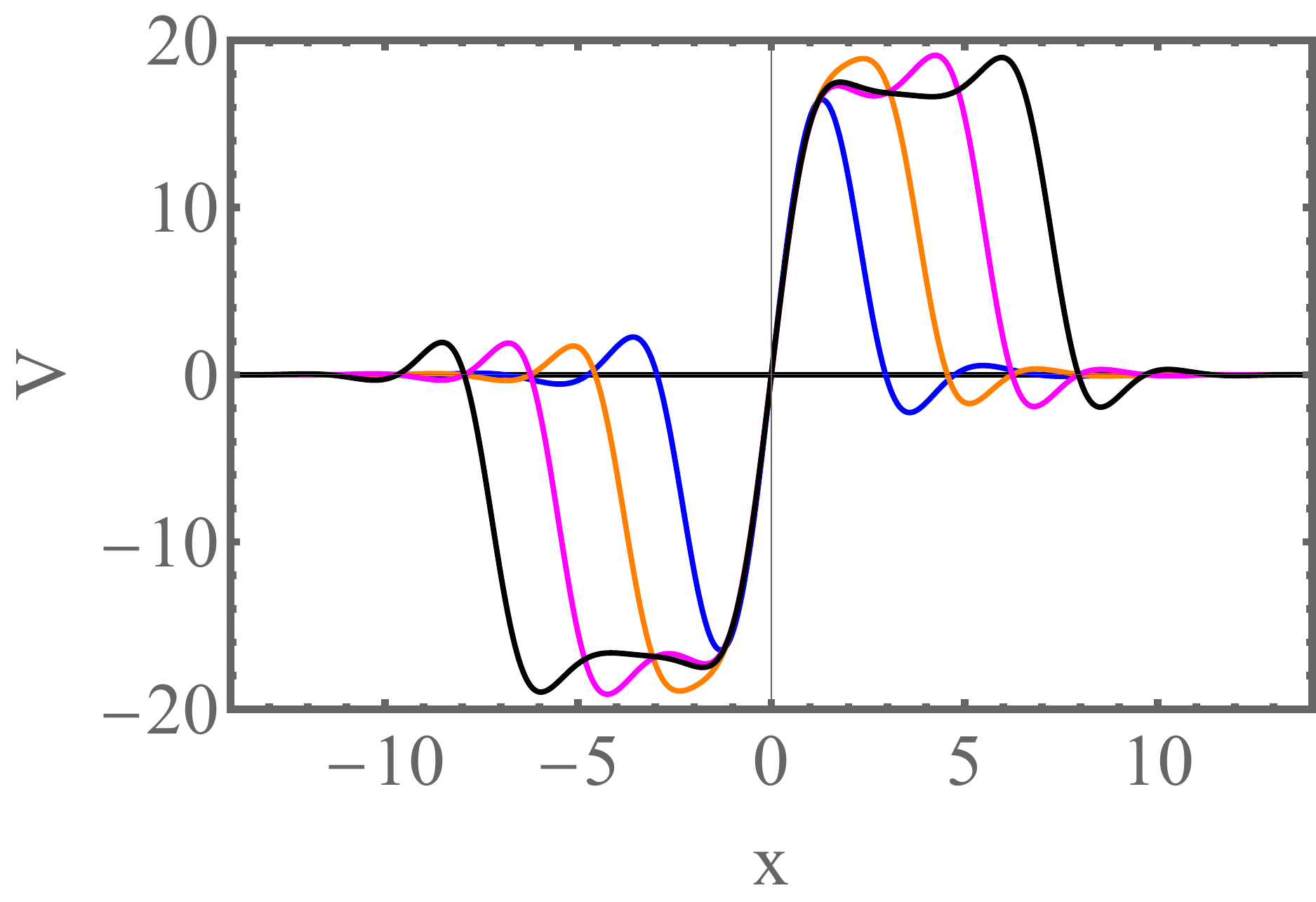}
\includegraphics[width=0.30\textwidth,clip=]{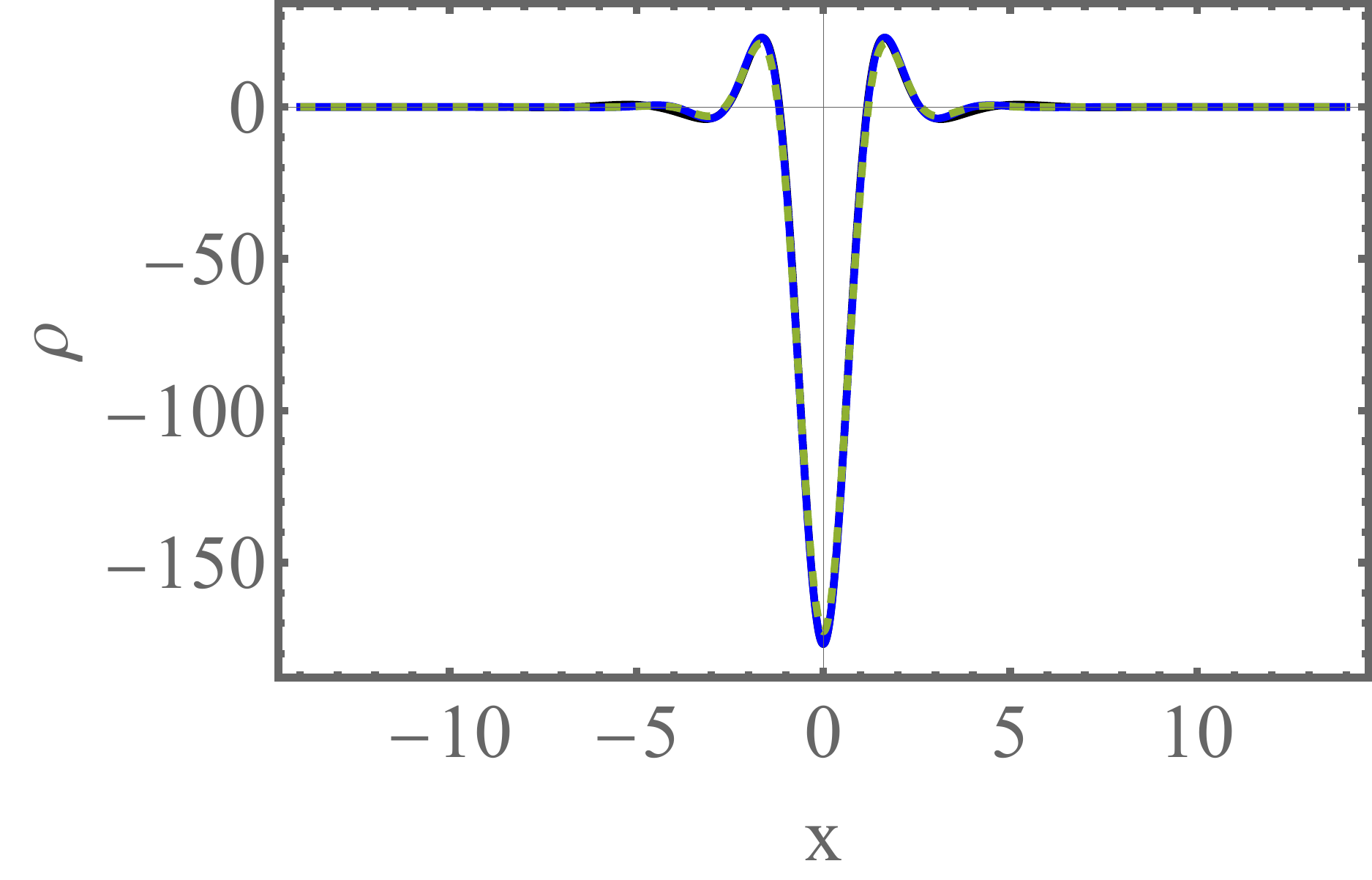}
\caption{The optimal path for large negative $\lambda H$. Shown are $h(x,t)$ (top panel) and $V(x,t)$ (middle panel) obtained by solving numerically  Eqs. (\ref{eqh}) and (\ref{eqrho}) for $\Lambda=-200$ (which corresponds to $H\simeq -143.6$) at times $t=0$, $0.25$, $0.5$, $0.75$ and $1$. The bottom panel shows $\rho(x,t)$ at times $t=0.3$, $0.6$ and $0.9$. The $V$-antishock, driven by a stationary $\rho$-soliton, and two outgoing deterministic $V$-shocks are clearly seen.}
\label{figCSnum}
\end{figure}

Similarly to the KPZ equation \cite{MKV}, the dominant contribution to the action in this tail
comes from a one-parameter family of solutions of the type
\begin{equation}\label{ansatz}
h(x,t)=h_0(x)-ct\quad \text{and} \quad \rho(x,t)=\rho_0(x)\,,
\end{equation}
parametrized by $c$. In this solution the (rescaled) optimal realization of the noise $\rho(x,t)$ represents a stationary soliton of $\rho_0(x)$ which drives a traveling wave of $h(x,t)$ in the vertical direction with velocity $c$ or, in the language of $V$,  a stationary antishock $V(x,t)=V_0(x)\equiv h_0^{\prime}(x)=0$ located at $x=0$. In contrast to the KPZ equation, here the soliton-antishock solution exhibits decaying spatial oscillations. As $c\gg 1$ (indeed, in the leading order $c=|H|\gg 1$), the soliton-antishock solution is strongly localized at $x=0$.  The $V$-antishock does not satisfy the boundary conditions $V(|x|\to \infty)=0$. Similarly to the KPZ equation, the remedy comes from two outgoing deterministic (that is, $\rho=0$) shocks of $V(x,t)$ which obey the equation
\begin{equation}
 \partial_{t} V +V \partial_x V = -\partial_{x}^4 V   \label{eqVdet}\\
\end{equation}
and satisfy the boundary conditions $V(x\to -\infty,t)=2c$ and $V(x\to \infty,t)=0$ for the right-moving shock, and $V(x\to -\infty,t)=0$ and $V(x\to \infty,t)=2c$  for the left-moving one.

We now present these solutions in some detail. Let us start with the soliton-antishock solution. Plugging the ansatz (\ref{ansatz}) into Eqs.~(\ref{eqh}) and (\ref{eqrho}) and integrating one of the two resulting ordinary differential equations with respect to $x$, we obtain
\begin{eqnarray}
  \frac{d^3V_0}{dx^3} &=& c-\frac{V_0^2}{2}+\rho_0\,,\label{VTW} \\
  \frac{d^3\rho_0}{dx^3}&=& \rho_0 V_0\,, \label{rhoTW}
\end{eqnarray}
where $V_0(x) = h_0^{\prime}(x)$, and the integration constant in Eq.~(\ref{rhoTW}) vanishes because $\rho(|x|\to \infty,t)=0$.
Upon rescaling
\begin{equation}\label{rescalingsoliton}
z=(c/2)^{1/6} x\,,\quad v=(2c)^{-1/2} V_0\,,\quad r=(1/c) \rho_0
\end{equation}
we recast Eqs.~(\ref{VTW}) and (\ref{rhoTW}) into a parameter-free form
\begin{eqnarray}
  v^{\prime\prime\prime} (z)&=& 1-v^2(z)+r(z)\,,\label{eqv} \\
  r^{\prime\prime\prime}(z) &=& 2 r(z) v(z)\,, \label{eqr}
\end{eqnarray}
where the primes denote the $z$-derivatives. Remarkably, the rescaling transformation (\ref{rescalingsoliton}) already makes it possible
to evaluate the rescaled action (\ref{action}) for this tail  up to a numerical constant $O(1)$ (which we will ultimately compute as well).  Indeed, using Eqs.~(\ref{action}) and (\ref{rescalingsoliton}), we obtain
\begin{equation}\label{supper}
  s \simeq \frac{1}{2} \int_{-\infty}^{\infty} dx \,\rho_0^2(x) = \beta c^{11/6} \simeq \beta |H|^{11/6}\,,
\end{equation}
where
\begin{equation}
\beta = \frac{1}{2^{5/6}} \int_{-\infty}^{\infty}  dz\, r^2(z)\,,
\label{beta}
\end{equation}
and we used the asymptotic relation $c\simeq |H|$.

\begin{figure}
\includegraphics[width=0.30\textwidth,clip=]{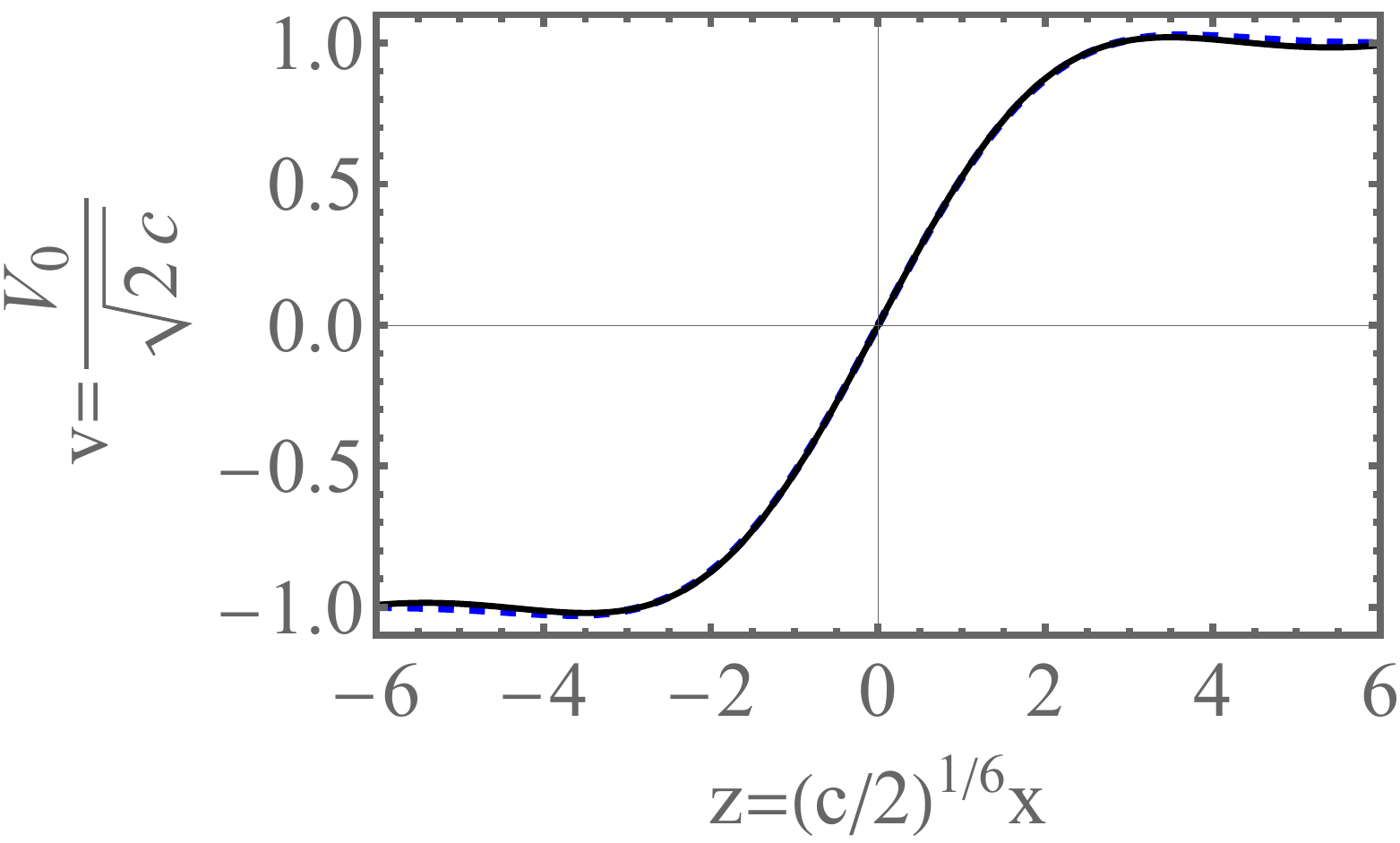}
\includegraphics[width=0.30\textwidth,clip=]{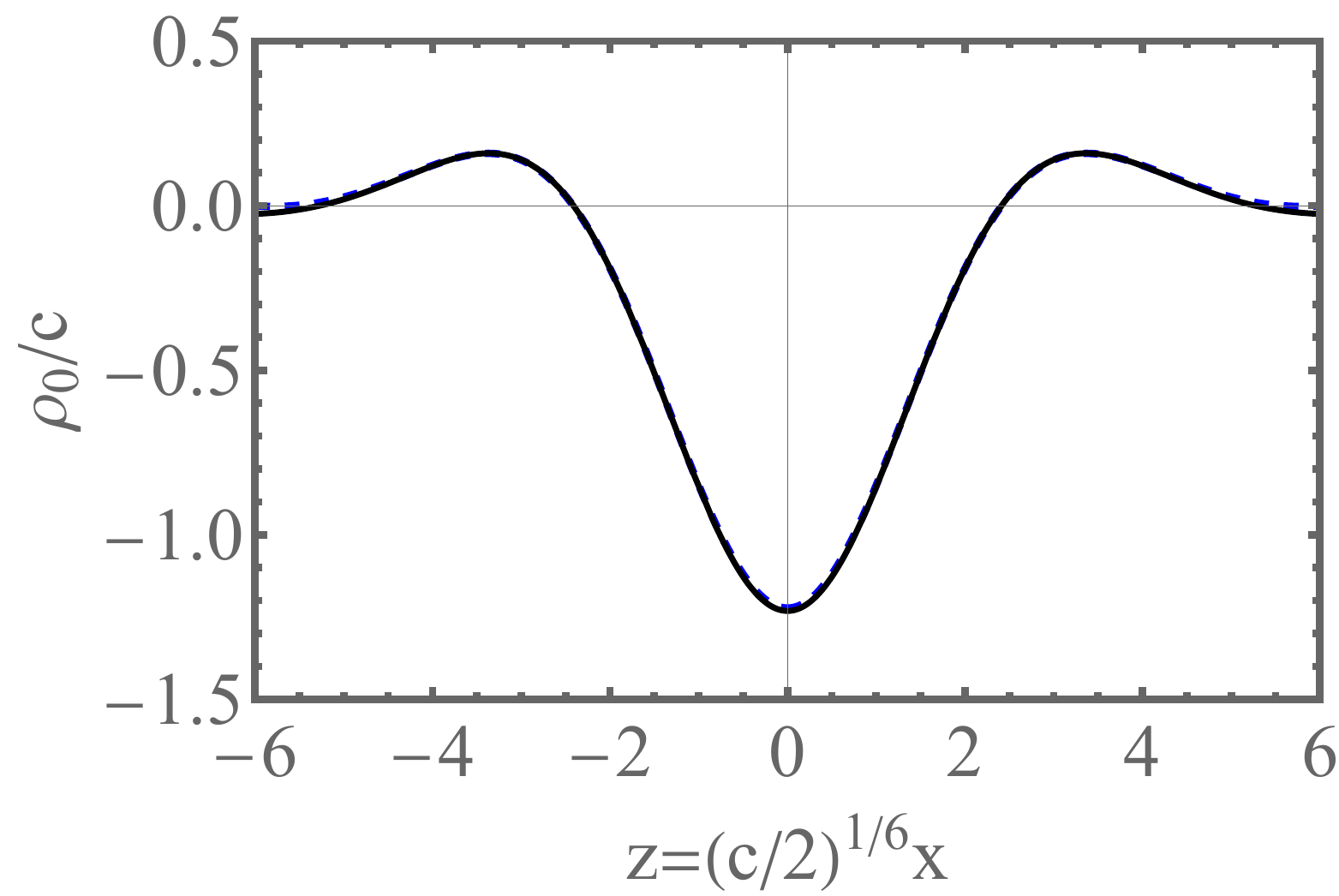}
\caption{The stationary soliton-antishock solution which dominates the upper tail $\lambda H\gg 1$ of $\mathcal{P}(H,T)$.  The dashed lines show a numerical solution of the ordinary equations~(\ref{eqv}) and (\ref{eqrho}) for the rescaled profiles $v(z)$ and $r(z)$. The solid lines show rescaled profiles $V(x,t=0.5)$ and $\rho(x,t=0.5)$, obtained by numerically solving the complete OFM equations~(\ref{eqV}) and (\ref{eqrhoV}) for $\Lambda=-200$.}
\label{figsoliton}
\end{figure}

To compute the constant $\beta =O(1)$, we have to solve the ordinary differential equations~(\ref{eqv}) and (\ref{eqr}). Analytical solution does not seem to be possible, so we resort to numerics. As $v(z)$ is anti-symmetric, and $r(z)$ symmetric, with respect to $z \leftrightarrow -z$, we can solve Eqs.~(\ref{eqv}) and (\ref{eqr}) only for $z>0$ with five boundary conditions $v(0)=v^{\prime\prime}(0)=r^{\prime}(0)=0$, $v(\infty)=1$ and $r(\infty)=0$. The resulting numerical solution for $v(z)$ and $r(z)$ is shown by dashed lines in Fig. \ref{figsoliton}. The functions $v(z)$ and $r(z)$ exhibit decaying oscillations. These can be easily understood from a linearization of Eqs.~(\ref{eqv}) and (\ref{eqr}) around the asymptotic states $v=1, r=0$ at $z\to \infty$ and $v=-1$ and $r=0$ at $z\to -\infty$.  Finally, using the numerically found $r(z)$, we obtain from Eq.~(\ref{beta}) $\beta\simeq 1.73$.

The same Fig. \ref{figsoliton} also shows, by solid lines, the $V$- and $\rho$-profiles obtained by  solving the full OFM equations (\ref{eqV}) and (\ref{eqrhoV}) numerically for $\Lambda=-200$.   The profiles were rescaled according to Eq.~(\ref{rescalingsoliton}), with $c=|H|\simeq 143.6$. As one can see, the agreement is good. Because of the boundary conditions in time, Eqs.~(\ref{H}) and (\ref{pT}), the soliton-antishock solution breaks down at $t$ close to $0$ and $1$. These narrow ``boundary layers" in time, however,  do not contribute to the action $s$ in the leading order in $|H|\gg 1$.

Now we turn to the outgoing deterministic shock solutions which take care of the boundary conditions at large distances. These shocks do not contribute to the action, and they are described by travelling front solutions of the form $V(x,t)=F(x\pm c_0 t)$ of Eq.~(\ref{eqVdet}). Equation~(\ref{eqVdet}) is a higher-order cousin of the Burgers equation $\partial_t V+V \partial_x V = \partial_x^2 V$. Because of the fourth-derivative dissipation term $-\partial_x^4 V$ the traveling shock of Eq.~(\ref{eqVdet}) exhibits decaying oscillations in space.  However, exactly as in the case of the Burgers equation, the shock velocity $c_0$ is independent of the dissipation, and it is equal to $(V_{+} + V_{-})/2$, where $V_+$ and $V_-$ are the asymptotic values of $V$ in front of and behind the shock, respectively.  For our right-moving shock $V_+ =0$ and $V_-=\sqrt{2c}$, and vice versa for the left-moving shock. Therefore, the deterministic shock velocity $c_0=\sqrt{c/2}\simeq \sqrt{|H|/2}$ is the same as for the KPZ equation \cite{MKV,Hopf}.

\begin{figure}
\includegraphics[width=0.30\textwidth,clip=]{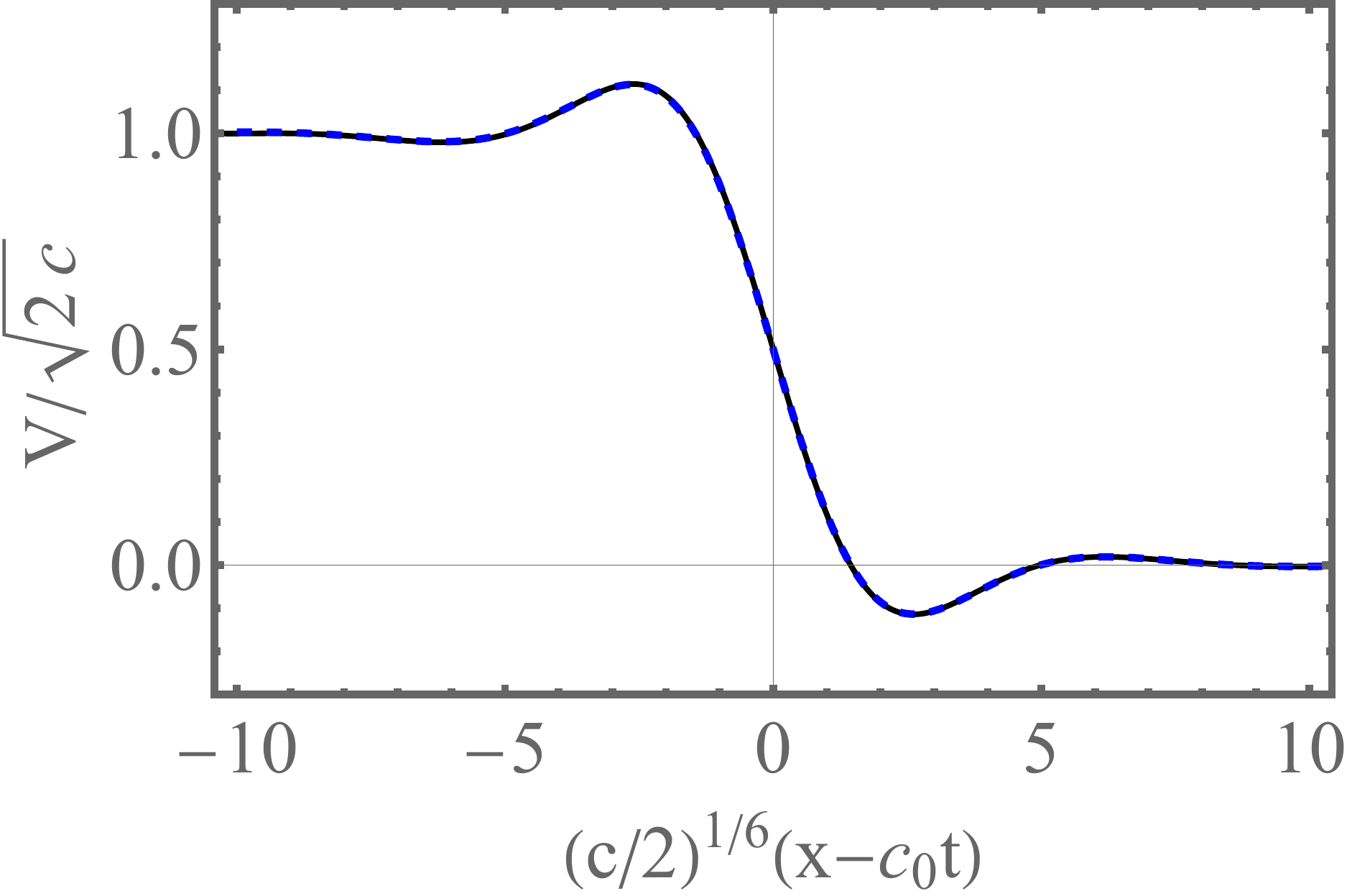}
\caption{Deterministic $V$-shock: an intrinsic feature of the optimal path dominating the upper tail $\lambda H\gg 1$ of $\mathcal{P}(H,T)$. Dashed line: travelling front solution $V(x-c_0t)$ of Eq.~(\ref{eqVdet}). When properly rescaled, this solution obeys Eq.~(\ref{detshockeq}). Solid line: rescaled profile $V(x,t=0.9)$, obtained by numerically solving the complete OFM problem [Eqs. (\ref{eqV}) and (\ref{eqrhoV})] for $\Lambda=-200$. Here $c=|H|\simeq 143.6$ and $c_0=\sqrt{|H|/2}\simeq 8.47$.}
\label{figplainshock}
\end{figure}

The shock profiles, however, are different. In particular, they approach their respective constant values at $x=\pm \infty$ in an oscillatory manner. To compute the profiles we rescale $v=V/\sqrt{2c}$ and $z=(x/2)^{1/6} (x-c_0t)$ and solve the resulting parameter-free ordinary differential equation
\begin{equation}\label{detshockeq}
v^{\prime\prime\prime}(z) = v(z)-v^2(z)
\end{equation}
numerically. Since $v(z)-1/2$ is an odd function of $z$, Eq.~(\ref{detshockeq}) can be solved only for $z>0$. For the right-moving shock the boundary conditions are $v(0)=1/2$, $v^{\prime\prime}(0)=0$ and $v(\infty)=0$. The solution  is shown by the dashed line in
Fig. \ref{figplainshock}. The same figure shows, by the solid line, the properly rescaled $V$-profiles obtained by  solving the full OFM equations (\ref{eqV}) and (\ref{eqrhoV}) for $\Lambda=-200$. A good agreement is observed.

\begin{figure}
\includegraphics[width=0.30\textwidth,clip=]{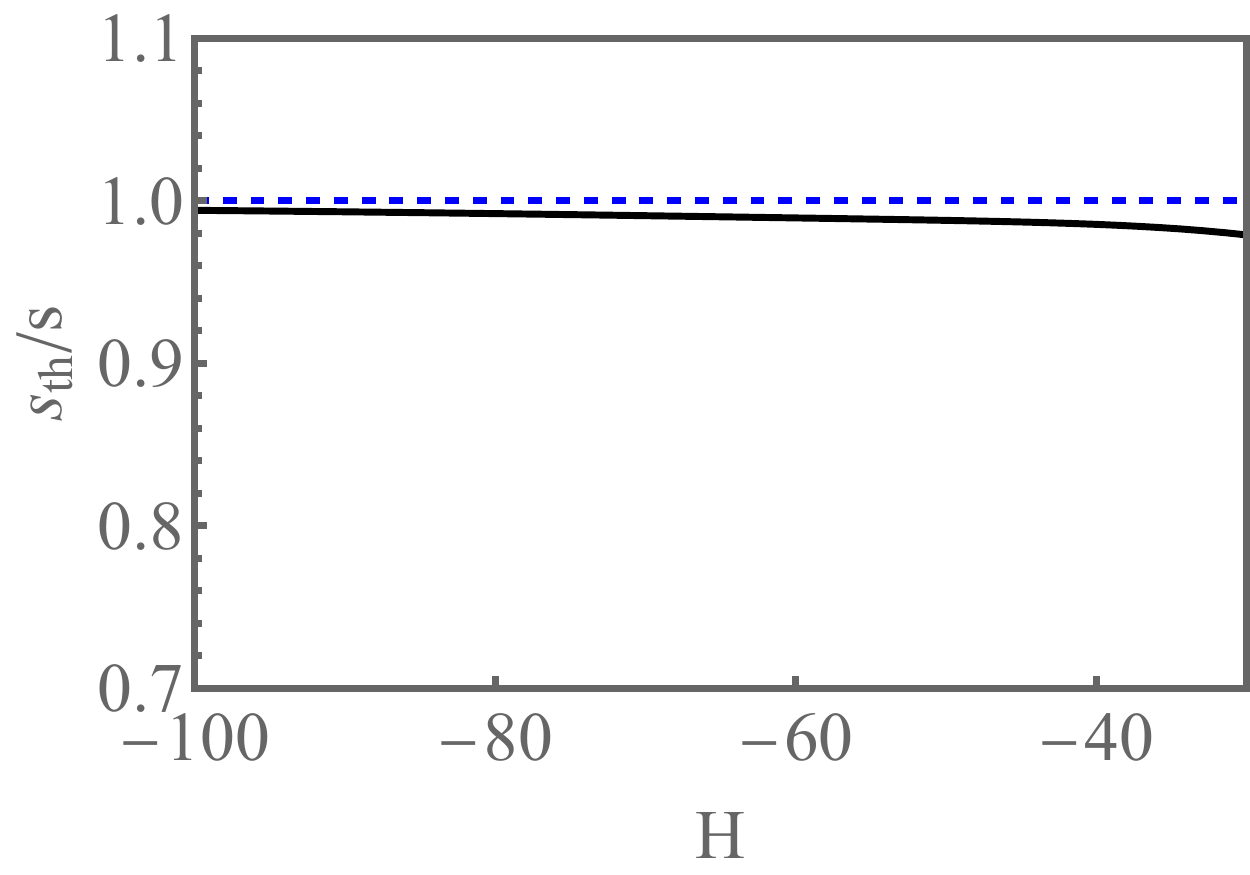}
\includegraphics[width=0.30\textwidth,clip=]{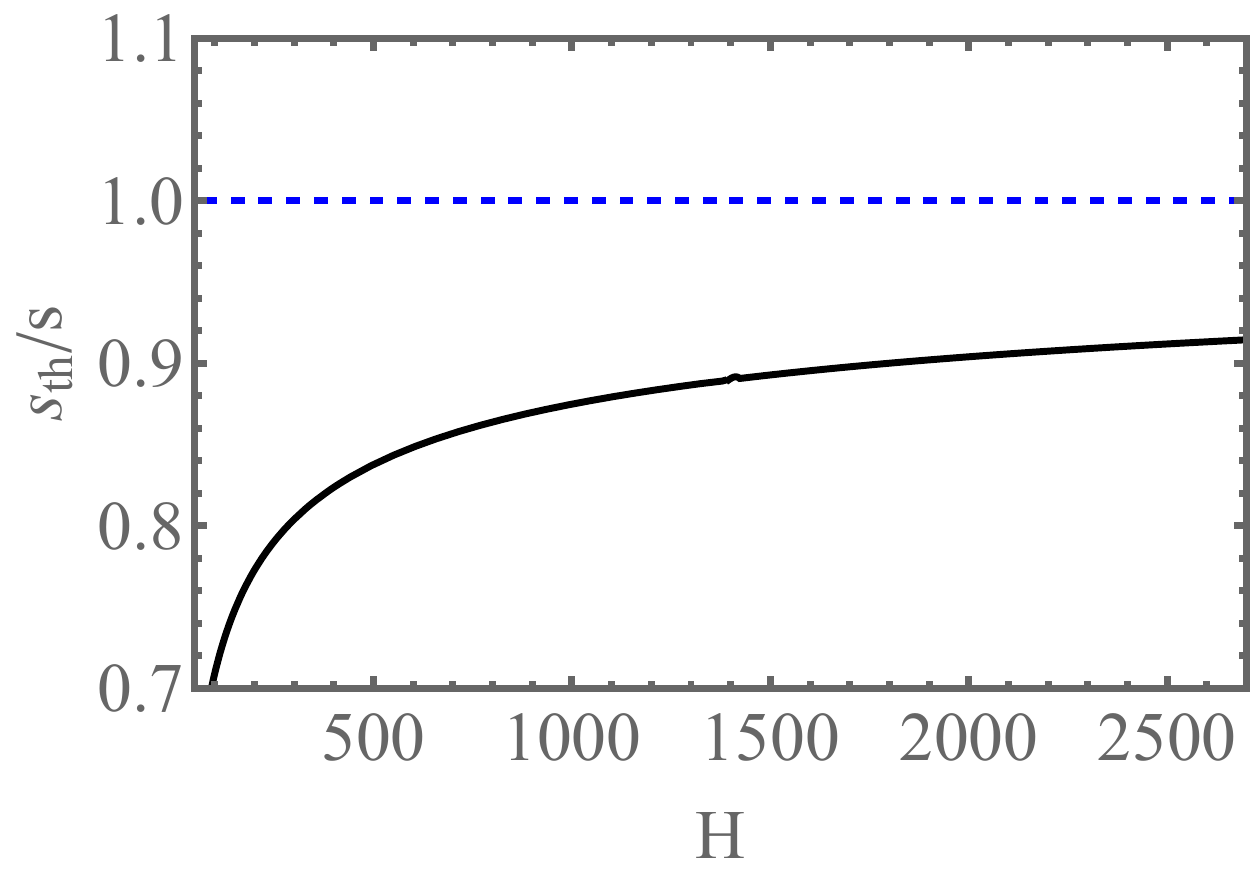}
\caption{Convergence of the asymptotics (\ref{supper}) (top) and (\ref{sHD}) (bottom) of the large-deviation function $s(H)$ to numerical results at large $|H|$.}
\label{figconvergence}
\end{figure}

The upper panel of Fig.~\ref{figconvergence} show that the asymptotic~(\ref{supper}) quickly converges to the numerical results at large negative $H$.
In the original variables Eq.~(\ref{supper}) yields the upper tail (\ref{11over6}) of $\mathcal{P}(H,T)$.

\subsection{Lower tail: $\lambda H \to -\infty$}
\begin{figure}
\includegraphics[width=0.30\textwidth,clip=]{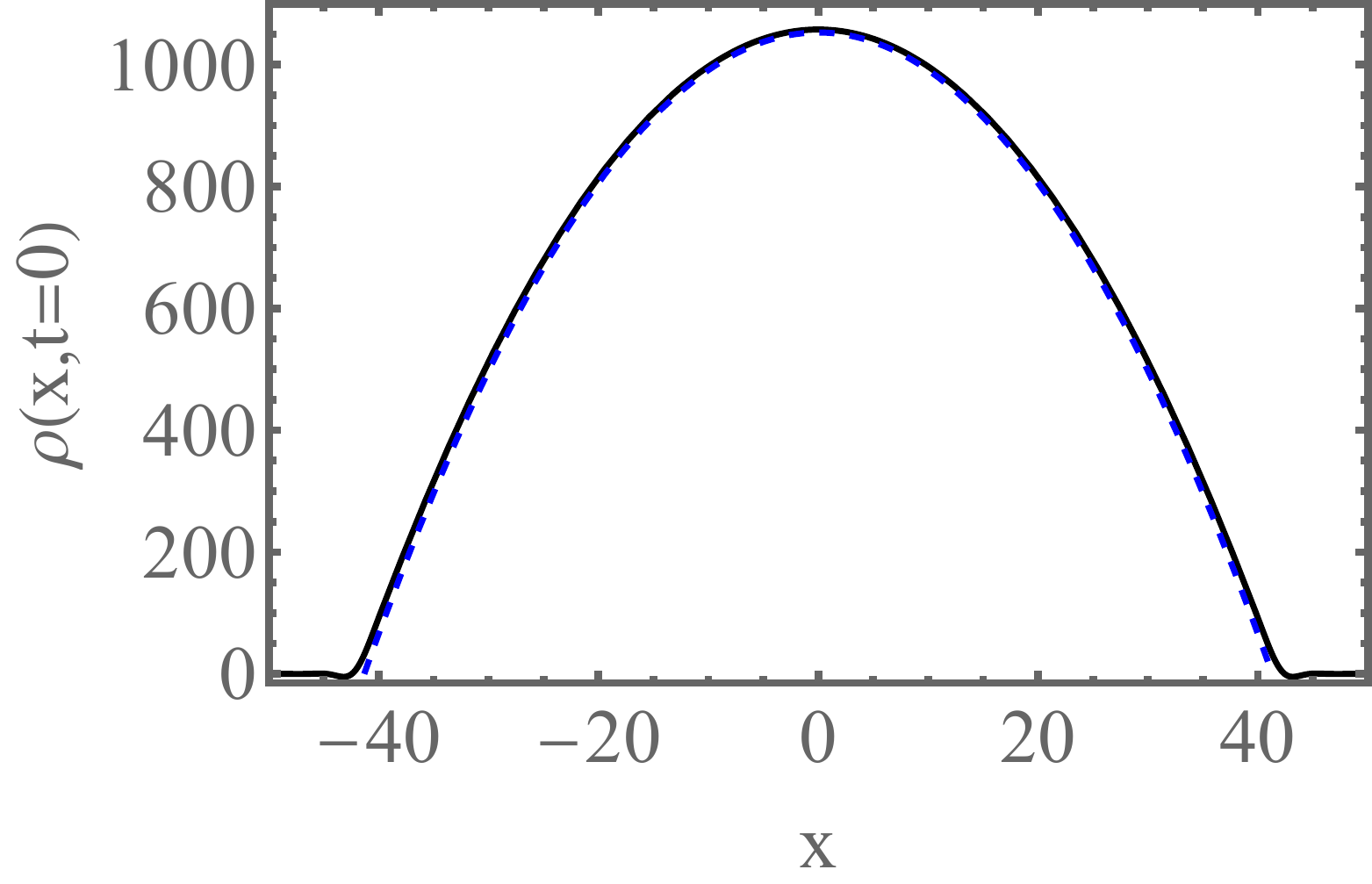}
\caption{Optimal configuration of the noise at $t=0$ which dominates  the lower tail $-\lambda H \gg 1$. Shown are the profiles of $\rho(x,t=0)$, obtained numerically here (the solid line) and analytically in Ref. \cite{MKV} (the dashed line), for $\Lambda=58 000$.}
\label{rhoHD}
\end{figure}

The lower tail is very different.  Similarly to the KPZ equation \cite{MKV}, the corresponding optimal path is described  by the zero-dissipation limit of Eqs.~(\ref{eqV}) and~(\ref{eqrhoV}):
\begin{eqnarray}
 \partial_t \rho +\partial_x (\rho V)&=& 0, \label{rhoeq}\\
  \partial_t V +V \partial_x V &=&\partial_x \rho\,. \label{Veq}
\end{eqnarray}
These hydrodynamic equations describe an inviscid flow of an effective compressible gas with density $\rho$ and velocity $V$, driven by the gradient of a \emph{negative} pressure $P(\rho)=-\rho^2/2$. The negative pressure causes collapse of the whole gas into the origin at $t=1$, in compliance with the boundary condition (\ref{pT}). This idealized problem is exactly soluble, and the solution is presented in  Ref. \cite{MKV}, see also Ref. \cite{KK2009}.  The spatial profile of $\rho(x,t)$ is parabolic in $x$ at all times, and it lives on a compact support which shrinks to zero at $t=1$. The spatial profile of $V(x,t)$ includes two regions. In the internal region, which has the same compact support as $\rho(x,t)$, the $V$-profile is linear in $x$. In the external region $V(x,t)$ comes from the solution of the Hopf equation $\partial_t V +V \partial_x V=0$ and its matching with the internal solution \cite{MKV}. This solution agrees very well with our numerical solution of the full system of equations (\ref{eqV}) and (\ref{eqrhoV}). As an example, Fig. \ref{rhoHD} shows $\rho(x,t=0)$, obtained numerically for $\Lambda=58,000$, and the corresponding theoretical profile of $\rho(x,t=0)$ obtained in Ref. \cite{MKV}.

The final result for the rescaled action $s(H)$ in this regime is \cite{KK2009,MKV}
\begin{equation}\label{sHD}
s(H)= \frac{8\sqrt{2}}{15\pi} H^{5/2}
\end{equation}
which, in the original variables, leads to the announced lower tail ~(\ref{5over2}) of $\mathcal{P}(H,T)$. The lower panel of Fig. \ref{figconvergence} shows that the asymptotic~(\ref{sHD}) slowly converges to the numerical result at large positive $H$. The slow convergence signals the presence of a relatively large subleading term in Eq.~(\ref{sHD}) which is not captured by the leading-order hydrodynamic solution. Our numerical data is compatible with a subleading term $O(H^2 \ln H)$.

The universal character of the leading-order tail asymptotic (\ref{sHD}) is explained by the macroscopic character of the flow which constitutes the optimal path in this regime. On large scales one can neglect dissipation of any origin if the dissipation is described by a differential operator of a higher than first order: be it diffusion, surface diffusion, \textit{etc.} As a result, for any stochastic interface model with the KPZ-type nonlinearity and non-conserved noise the lower tail of $\mathcal{P}(H,T)$ at short times and in one dimension will be described by Eq.~(\ref{5over2}).

\section{Summary and Discussion}
\label{discussion}

We studied large deviations of the one-point height distribution $\mathcal{P}(H,T)$ of an initially flat interface, governed by the GB equation (\ref{GBdimensional}). We focused on the short-time limit. Employing the OFM, we addressed both typical fluctuations, see Eq.~(\ref{Gauss}), and the distribution tails (\ref{11over6}) and (\ref{5over2}).  The upper tail (\ref{11over6}) turns out to be very different from the upper tail of the KPZ equation. The lower tail (\ref{5over2}) coincides with its counterpart for the KPZ equation, and the reason for that is a macroscopic character of the optimal path and, as a consequence, its insensitivity to any dissipation mechanism which is described by a differential operator of a higher than first order.

Importantly, the optimal paths of the interface in this problem respect, for all $H$, the mirror symmetry, $h(-x,t)=h(x,t)$, $V(-x,t) = - V(x,t)$, and $\rho(-x,t)=\rho(x,t)$. There are two important consequences of this fact.  The first is that the upper tail $\lambda H \to \infty$ of $\mathcal{P}(H,T)$, see Eq.~(\ref{11over6}), is universal for a whole class of deterministic initial conditions. The reason is that the $\rho$-soliton and $V$-antishock are strongly localized at $x=0$, while two expanding deterministic shocks make it possible to match the small soliton-antishock region with macroscopic external regions. The latter accommodate the problem-specific boundary conditions, but do not contribute to the action in the leading order.

The second consequence of the mirror symmetry of the optimal paths comes in the form of a simple connection between the action $s(H)$, corresponding to $\mathcal{P}(H,T)$  for the infinite interface $|x|<\infty$ and the action $s_{1/2}(H)$, corresponding to $\mathcal{P}(H,T)$ for the half-infinite interface $0\leq x<\infty$ with a reflecting wall at $x=0$. Indeed,  exploiting the mirror symmetry,  we obtain
\begin{equation}\label{actionhalf}
s_{1/2}(H) = \frac{1}{2}\int_0^1 dt \int_{0}^{\infty} dx\,\rho^2(x,t) = \frac{s(H)}{2}\,.
\end{equation}
That is, the probability of observing a specified value of $H$ in the half-infinite system is always higher than in the infinite system, and it is \emph{exponentially} higher in the tails. The same property is observed for the KPZ equation, for all deterministic initial conditions that respect the mirror symmetry \cite{SMS2018}.

It would be also interesting to evaluate the short-time distribution $\mathcal{P}(H,T)$ in higher dimensions. In $d$ dimensions the GB-equation, rescaled according to Eq. (\ref{rescaling}), acquires  the following form:
\begin{equation}\label{GBd}
\partial_t h=-\nu \nabla^4 h -\frac{1}{2}(\nabla h)^2 +\sqrt{\epsilon_d}\, \xi(\mathbf{x},t)\,,
\end{equation}
where we have again assumed, without loss of generality, that $\lambda<0$. In Eq.~(\ref{GBd})
\begin{equation}\label{epsilond}
\epsilon_d =\frac{D T^\frac{8-d}{4} \lambda^2}{\nu^{\frac{d+4}{4}}}
= \left(\frac{T}{T_{\text{NL}}}\right)^{\frac{8-d}{4}}\,,
\end{equation}
where
\begin{equation}\label{TNLd}
T_{\text{NL}}=\frac{\nu^{\frac{d+4}{8-d}}}{D^{\frac{4}{8-d}}|\lambda|^{\frac{8}{8-d}}}
\end{equation}
is the characteristic nonlinear time. It is clear from Eq.~(\ref{epsilond}) that,  at $d<d_c=8$, the parameter $\epsilon_{d}$ is small at short times. A more stringent technical limitation comes from fact that  the short-time variance of $\mathcal{P}(H,T)$ -- which is determined by the $d$-dimensional version of the linear equation~(\ref{MH}) --  is well-defined (and therefore does not require a regularization by a small-scale cutoff  or by local averaging) only at $d<4$ \cite{Krug1997,SMS2017}.  In any case, in all physical dimensions,  the short-time height statistics $\mathcal{P}(H,T)$ of the GB interface (i) is well defined, and (ii) it can be captured by the OFM \cite{incontrast}.

\vspace{0.5cm}
\section*{Acknowledgement}
BM acknowledges support from the Israel Science Foundation (ISF) through Grant No. 1499/20.


\begin{thebibliography}{99}
\bibitem{numerics} We implemented a modified version of the back-and-forth iteration algorithm \cite{Chernykh} as described in Ref. \cite{MVK}. Some of the computations were done using standard tools of ``Mathematica".
\bibitem{Hopf} Since $c_0 \simeq \sqrt{|H|/2}\gg 1$, the characteristic width of the outgoing shocks is much smaller than the distance $\simeq |H|$ each of them travels until $t=1$. Therefore, at large scales the shocks can be described by the zero-dissipation Hopf equation $\partial_tV+V\partial_x V=0$.
\end{thebibliography}
\end{document}